\documentclass[
    ,final            
  ,bibliocites                 
  ]
  {aipproc}

\layoutstyle{6x9}

\newcommand{\apj}{ApJ}
\newcommand{\apjl}{ApJ}
\newcommand{\apjs}{ApJS}

\newcommand{\aap}{A\&A}

\begin{document}

\title{Homogeneous comparison of planet candidates imaged directly until 2008}


\author{Tobias O. B. Schmidt}{
  address={Astrophysikalisches Institut, Universit\"at Jena, Schillerg\"asschen 2-3, 07745 Jena, Germany, tobi@astro.uni-jena.de}
}

\author{Ralph Neuh\"auser}{
  address={Astrophysikalisches Institut, Universit\"at Jena, Schillerg\"asschen 2-3, 07745 Jena, Germany}
}

\author{Andreas Seifahrt}{
  address={Institut f\"ur Astrophysik, Universit\"at G\"ottingen, Friedrich-Hund-Platz 1, 37077 G\"ottingen, Germany}
}

\begin{abstract}
We present a compilation of the planet candidates currently known from direct imaging. We have gathered available data from the literature and derive the luminosity of all candidates in a homogeneous way using a bolometric correction, the distances and the K band magnitudes of the objects. In a final step we find the masses of the candidates from a comparison of the luminosity or, if not available, an absolute brightness and several well known hot-start evolutionary models.
\end{abstract}

\maketitle


\section{Introduction}

Searching the literature, we could find 20 directly imaged planet candidates. We gathered photometry as well as age information of these objects (see Table 1) and derived their luminosity in a homogeneous way, taking a bolometric correction into account. We finally use the information to find object masses from several well known hot-start evolutionary models (see Table 2) and present an overview plot in Figure 1.

\section{Discussion}

While done for most objects, the common proper motion confirmation is still needed for some objects, as e.g. 1RXS J1609, $\beta$ Pic, or CHXR 73. Since there is no clear definition of planet outside of the solar system (13 Jupiter masses or below brown dwarf desert), yet, it is moreover unclear if planets can also occur around brown dwarfs, as it would be the case for e.g.~2M1207, UScoCTIO 108 or FU Tau. Hence, the companion of Fomalhaut is currently the most conclusive candidate for a planet found by direct imaging with 3 M$_{Jup}$ upper mass limit. The Burrows et al.~(1997) model works best on 2M0535 B (37.7 $\pm$ 2.9 M$_{Jup}$ dynamic calibration), the Wuchterl model seems worst.




\begin{table}
\centering
\begin{tabular}{cccccl}
\hline
Object    & Luminosity                     & Magnitude & Temperature & Age  & References \\
name      & $\log (L_{\rm bol}/L_{\odot})$ & M$_{\rm K}$ [mag] & T$_{\rm eff}$ [K] & [Myrs] & \\ \hline
\multicolumn{6}{c}{Reference object (eSB2 brown dwarf - brown dwarf binary 2M0535):} \\ 
\hline
2M0535 A  & $-1.65 \pm 0.07$ & $5.29 \pm 0.16$ & $2715 \pm 100$ & 1 (0-3) & Stassun 2007 \\
       B  & $-1.83 \pm 0.07$ & $5.29 \pm 0.16$ & $2820 \pm 105$ & 1 (0-3) & Stassun 2007 \\
\hline
\multicolumn{6}{c}{Directly detected planet candidates:} \\
\hline
DH Tau b      & $-2.44 $         &                  & $2750 \pm 50$  & 10      & Itoh 2005 \\
              & $-2.75 \pm 0.10$ & $8.31 \pm 0.23$  &                & 0.1-4   & [17], [22] \\
GQ Lup b      & $-2.38 \pm 0.25$ & $7.67 \pm 0.16$  & $2650 \pm 100$ & 1-3     & Neuh. 05, [32], [34] \\
2M1207 A      & $-2.76 \pm 0.05$ & $8.35 \pm 0.05$  & $2425 \pm 160$ & 5-12    & Chauvin 05a, [11] \\
       b      & $-4.75 \pm 0.06$ & $13.33 \pm 0.12$ & $1590 \pm 280$ & 5-12    & [7], [16], [26], [31] \\
AB Pic b      & $-3.76 \pm 0.06$ & $10.85 \pm 0.11$ & $2040 \pm 160$ & 20-40   & Chauvin 05b, [31]\\
Oph 1622 a    & $-2.61 \pm 0.23$ & $7.96 \pm 0.56$  &                & 5 (1-20)& Allers 2006, [2] \\
Oph 1622 b    & $-2.79 \pm 0.23$ & $8.42 \pm 0.56$  &                & 5 (1-20)& [8], [15]. [18]. [24] \\
SCR 1845 b    & \multicolumn{2}{c}{M$_{\rm H} = 15.30\,+0.31\,-0.26$}& (850)   & 100-10$^{4}$& Biller 2006 \\
CHXR 73 b     & $-2.62 \pm 0.21$ & $8.00 \pm 0.52$  & $2600 \pm 450$ & 0.1-4   & Luhman 2006, [33] \\
HD 203030 b   & $-4.69 \pm 0.07$ & $13.15 \pm 0.14$ & late L         & 130-400 & Metchev 2006 \\
HN Peg b      & $-4.94 \pm 0.04$ & $13.80 \pm 0.05$ & early T        & 100-500 & Luhman 2007 \\
USco 108 b    & $-3.14 \pm 0.14$ & $9.30 \pm 0.34$  & (2350)         & 5 (1-10)& B{\'e}jar 2008 \\
CT Cha b      & $-2.68 \pm 0.21$ & $8.83 \pm 0.50$  & $2600 \pm 250$ & 0.1-4   & Schmidt 2008 \\
1RXSJ1609 b   & $-3.57 \pm 0.15$ & $10.36 \pm 0.35$ & early L        & 5 (1-10)& Lafreni{\`e}re 2008 \\
Fomalhaut b   & $\le -6.5$       & M$_{\rm H} \ge 23.5$ &            & 100-300 & Kalas 2008 \\
HR 8799 b     & $-5.1 \pm 0.1$   & $14.05 \pm 0.08$ &                & 20-1128 & Marois 2008 \\
        c     & $-4.7 \pm 0.1$   & $13.13 \pm 0.08$ &                & 20-1128 & [14] \\
        d     & $-4.7 \pm 0.1$   & $13.11 \pm 0.12$ &                & 20-1128 & [35] \\
$\beta$ Pic b & \multicolumn{2}{c}{M$_{\rm L'} = 9.77 \pm 0.30$}   & & 8-20    & Lagrange 2009 \\
FU Tau b      & $-2.40 \pm 0.09$ & $7.44 \pm 0.20$  &                & 0.1-4   & Luhman 2009 \\
EK 60  b      & $-3.14 \pm 0.18$ & $9.28 \pm 0.44$  &                & 0.3-10  & Kuzuhara (in prep.)\\
\hline
\end{tabular}
\caption{Observed properties of the directly imaged planet candidates.}
\label{table:1}
\end{table}

\begin{table}
\begin{tabular}{ccccccc}
\hline
Object    & Burrows 97 & Chabrier 00 & Baraffe 03 & Marley 07 & Baraffe 08 & Wuchterl \\ 
name     & (L,\,age)     & (L,\,M$_{\rm K}$,\,age) & (L,\,M$_{\rm K}$,\,age) & ($\le10$\,Jup) & ($\ge10$\,Myrs)  & (Neuh05)\\ 
\hline
\multicolumn{7}{c}{Reference object (eSB2 brown dwarf - brown dwarf binary 2M0335):} \\ 
\hline
2M0535 A      & 50 (45-60)   & 55 (30-60)     & 50 (45-80)     &          &              & 5-13    \\
       B      & 37 (33-46)   & 45 (40-50)     & 43 (40-65)     &          &              & $\le 13$ \\ 
\hline 
\multicolumn{7}{c}{Directly detected planet candidates:} \\ 
\hline
DH Tau b      & 14 (12-19)   & 11 (10-20)     & 10 (7-20)      &          &              & 5 \\
GQ Lup b      & 19 (16-30)   & 22 (12-30)     & 17 (12-32)     &          &              & 1-5 \\
2M1207 A      & 18 (18-19)   & 20 (17-23)     & 19 (15-21)     &          &              & 1-5 \\
       b      & 4 (3-5)      & 4 (2-5)        & 3 (1-4)        & 4 (3-5)  &      4       & \\
AB Pic b      & 13.5 (13-14) & 15 (11-20)     & 13 (11-20)     &          &              & \\
Oph 1622 a    & 20 (14-55)   & 20 (10-40)     & 20 (9-40)      &          &              & \\
Oph 1622 b    & 18 (10-40)   & 17 (7-25)      & 17 (7-23)      &          &              & \\
SCR 1845 b    &              &  11.5-65       & 9-65           &          &              & \\
CHXR 73 b     & 17 (12-23)   & 13 (10-25)     & 12 (8-25)      &          &              & 2-5 \\
HD 203030 b   & 17 (12.5-30) & $\le 30$       & 18 (11-30)     &          &              & \\
HN Peg b      & 17 (11.5-25) & 17 (10-25)     & 17 (10-23)     & $\ge 10$ &              & \\
USco 108 b    & 14 (7-17)    & 13.5 (5-17)    & 12 (5-18)      &          &              & 1-2\\
CT Cha b      & 17 (11-23)   & 13 (8-25)      & 12 (6-30)      &          &              & 2-5 \\ 
1RXSJ1609 b   & 9.5 (4-14)   &  8 (3-13)      & 8 (3-12)       & 8 ($\ge 4$)&            & \\ 
Fomalhaut b   & $\le 4$      &                & $\le 3$        & $\le 3$  & $\le 2$      & \\ 
HR 8799 b     & 8 (4-36)     & $\le 35$       & 6 (4-38)       & 7 ($\ge 3$)& 7 ($\ge 3$)& \\
        c     & 12 (6-50)    & 10 (6-46)      & 9 (6-48)       & 10 ($\ge 6$)& $\ge 5$   & \\
        d     & 12 (6-50)    & 10 (6-46)      & 9 (6-48)       & 10 ($\ge 6$)& $\ge 5$   & \\ 
$\beta$ Pic b &              & 7.5 (6-10)     & 9 (6-12)       &          &              & \\ 
FU Tau b      & 18 ($\le 25$)& 20 (15-28)     & 17 (14-30)     &          &              & \\
EK 60  b      & 14 (6-17)    & 13.5 (5-17)    & 12.5 (5-17)    & $\ge 5$  &              & \\
\hline
\end{tabular}
\caption{Masses derived from evolutionary hot-start models.}
\label{table:2}
\end{table}

\begin{figure}
  \centering
  \includegraphics[height=.45\textheight]{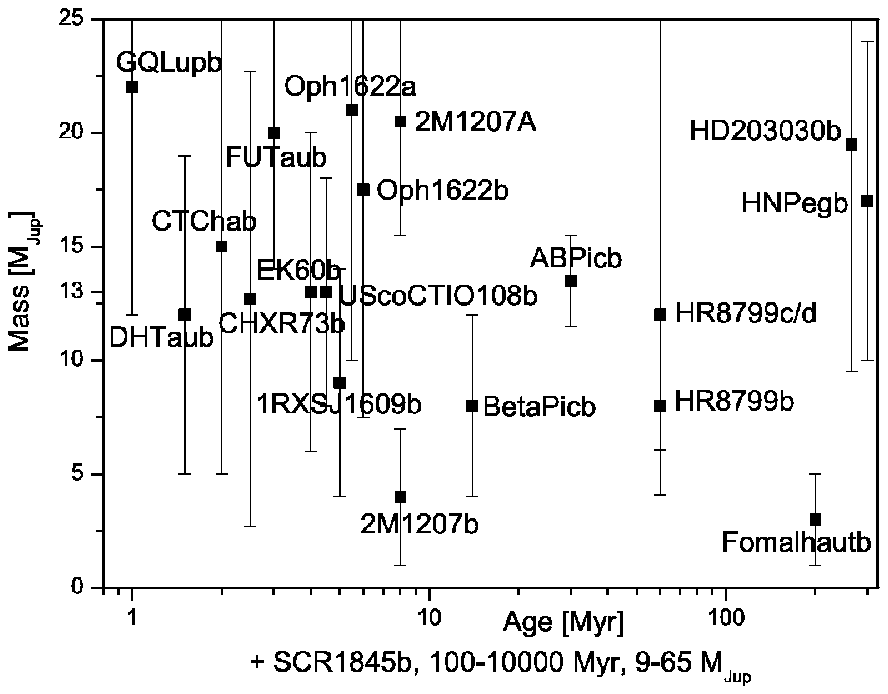}
  \caption{Derived masses of the planet candidates found by direct imaging (see Table 1 \& 2). For reasons of readability the error bars in age were not plotted and some of the objects were slightly moved from their repective best age estimate.}
\label{fig:1}
\end{figure}


\begin{theacknowledgments}
TOBS would like to thank Evangelisches Studienwerk e.V. Villigst for financial support. This research has made use of NASA's Astrophysics Data System Bibliographic Services and the Simbad database, operated at the Observatoire Strasbourg.
\end{theacknowledgments}



\bibliographystyle{aipprocl} 


\begin{thebibliography}{10}
\providecommand{\enquote}[1]{``#1''}
\expandafter\ifx\csname url\endcsname\relax
  \def\url#1{\texttt{#1}}\fi
\expandafter\ifx\csname urlprefix\endcsname\relax\def\urlprefix{URL }\fi

\bibitem{2006PhDT.........2A}
K.~N. {Allers}, \emph{{Disks and dissociation regions: The interaction of young
  stellar objects with their environments}}, Ph.D. thesis, The University of
  Texas at Austin, United States -- Texas (2006).

\bibitem{2007ApJ...657..511A}
K.~N. {Allers}, D.~T. {Jaffe}, K.~L. {Luhman}, M.~C. {Liu}, J.~C. {Wilson},
  M.~F. {Skrutskie}, M.~{Nelson}, D.~E. {Peterson}, J.~D. {Smith}, and M.~C.
  {Cushing}, \emph{\apj} \textbf{657}, 511--520 (2007).

\bibitem{2003AA...402..701B}
I.~{Baraffe}, G.~{Chabrier}, T.~S. {Barman}, F.~{Allard}, and P.~H.
  {Hauschildt}, \emph{\aap} \textbf{402}, 701--712 (2003).

\bibitem{2008AA...482..315B}
I.~{Baraffe}, G.~{Chabrier}, and T.~{Barman}, \emph{\aap} \textbf{482},
  315--332 (2008).

\bibitem{2008ApJ...673L.185B}
V.~J.~S. {B{\'e}jar}, M.~R. {Zapatero Osorio}, A.~{P{\'e}rez-Garrido},
  C.~{{\'A}lvarez}, E.~L. {Mart{\'{\i}}n}, R.~{Rebolo},
  I.~{Vill{\'o}-P{\'e}rez}, and A.~{D{\'{\i}}az-S{\'a}nchez}, \emph{\apjl}
  \textbf{673}, L185--L189 (2008).

\bibitem{2006ApJ...641L.141B}
B.~A. {Biller}, M.~{Kasper}, L.~M. {Close}, W.~{Brandner}, and S.~{Kellner},
  \emph{\apjl} \textbf{641}, L141--L144 (2006).

\bibitem{2007ApJ...669L..41B}
B.~A. {Biller}, and L.~M. {Close}, \emph{\apjl} \textbf{669}, L41--L44 (2007).

\bibitem{2006ApJ...653L..61B}
A.~{Brandeker}, R.~{Jayawardhana}, V.~D. {Ivanov}, and R.~{Kurtev},
  \emph{\apjl} \textbf{653}, L61--L64 (2006).

\bibitem{1997ApJ...491..856B}
A.~{Burrows}, M.~{Marley}, W.~B. {Hubbard}, J.~I. {Lunine}, T.~{Guillot},
  D.~{Saumon}, R.~{Freedman}, D.~{Sudarsky}, and C.~{Sharp}, \emph{\apj}
  \textbf{491}, 856--+ (1997).

\bibitem{2000ApJ...542..464C}
G.~{Chabrier}, I.~{Baraffe}, F.~{Allard}, and P.~{Hauschildt}, \emph{\apj}
  \textbf{542}, 464--472 (2000).

\bibitem{2004AA...425L..29C}
G.~{Chauvin}, A.-M. {Lagrange}, C.~{Dumas}, B.~{Zuckerman}, D.~{Mouillet},
  I.~{Song}, J.-L. {Beuzit}, and P.~{Lowrance}, \emph{\aap} \textbf{425},
  L29--L32 (2004).

\bibitem{2005AA...438L..25C}
G.~{Chauvin}, A.-M. {Lagrange}, C.~{Dumas}, B.~{Zuckerman}, D.~{Mouillet},
  I.~{Song}, J.-L. {Beuzit}, and P.~{Lowrance}, \emph{\aap} \textbf{438},
  L25--L28 (2005a).

\bibitem{2005AA...438L..29C}
G.~{Chauvin}, A.-M. {Lagrange}, B.~{Zuckerman}, C.~{Dumas}, D.~{Mouillet},
  I.~{Song}, J.-L. {Beuzit}, P.~{Lowrance}, and M.~S. {Bessell}, \emph{\aap}
  \textbf{438}, L29--L32 (2005b).

\bibitem{2006ApJS..166..351C}
C.~H. {Chen}, B.~A. {Sargent}, C.~{Bohac}, K.~H. {Kim}, E.~{Leibensperger},
  M.~{Jura}, J.~{Najita}, W.~J. {Forrest}, D.~M. {Watson}, G.~C. {Sloan}, and
  L.~D. {Keller}, \emph{\apjs} \textbf{166}, 351--377 (2006).

\bibitem{2007ApJ...660.1492C}
L.~M. {Close}, B.~{Zuckerman}, I.~{Song}, T.~{Barman}, C.~{Marois}, E.~L.
  {Rice}, N.~{Siegler}, B.~{Macintosh}, E.~E. {Becklin}, R.~{Campbell}, J.~E.
  {Lyke}, A.~{Conrad}, and D.~{Le Mignant}, \emph{\apj} \textbf{660},
  1492--1506 (2007).

\bibitem{2008AA...477L...1D}
C.~{Ducourant}, R.~{Teixeira}, G.~{Chauvin}, G.~{Daigne}, J.-F. {Le Campion},
  I.~{Song}, and B.~{Zuckerman}, \emph{\aap} \textbf{477}, L1--L4 (2008).

\bibitem{2005ApJ...620..984I}
Y.~{Itoh}, M.~{Hayashi}, M.~{Tamura}, T.~{Tsuji}, Y.~{Oasa}, M.~{Fukagawa},
  S.~S. {Hayashi}, T.~{Naoi}, M.~{Ishii}, S.~{Mayama}, J.-i. {Morino},
  T.~{Yamashita}, T.-S. {Pyo}, T.~{Nishikawa}, T.~{Usuda}, K.~{Murakawa},
  H.~{Suto}, S.~{Oya}, N.~{Takato}, H.~{Ando}, S.~M. {Miyama}, N.~{Kobayashi},
  and N.~{Kaifu}, \emph{\apj} \textbf{620}, 984--993 (2005).

\bibitem{2006Sci...313.1279J}
R.~{Jayawardhana}, and V.~D. {Ivanov}, \emph{Science} \textbf{313}, 1279--1281
  (2006).

\bibitem{2008Sci...322.1345K}
P.~{Kalas}, J.~R. {Graham}, E.~{Chiang}, M.~P. {Fitzgerald}, M.~{Clampin},
  E.~S. {Kite}, K.~{Stapelfeldt}, C.~{Marois}, and J.~{Krist}, \emph{Science}
  \textbf{322}, 1345-- (2008).

\bibitem{2008ApJ...689L.153L}
D.~{Lafreni{\`e}re}, R.~{Jayawardhana}, and M.~H. {van Kerkwijk}, \emph{\apjl}
  \textbf{689}, L153--L156 (2008).

\bibitem{2009AA...493L..21L}
A.-M. {Lagrange}, D.~{Gratadour}, G.~{Chauvin}, T.~{Fusco}, D.~{Ehrenreich},
  D.~{Mouillet}, G.~{Rousset}, D.~{Rouan}, F.~{Allard}, {\'E}.~{Gendron},
  J.~{Charton}, L.~{Mugnier}, P.~{Rabou}, J.~{Montri}, and F.~{Lacombe},
  \emph{\aap} \textbf{493}, L21--L25 (2009).

\bibitem{2006ApJ...649..894L}
K.~L. {Luhman}, J.~C. {Wilson}, W.~{Brandner}, M.~F. {Skrutskie}, M.~J.
  {Nelson}, J.~D. {Smith}, D.~E. {Peterson}, M.~C. {Cushing}, and E.~{Young},
  \emph{\apj} \textbf{649}, 894--899 (2006).

\bibitem{2007ApJ...654..570L}
K.~L. {Luhman}, B.~M. {Patten}, M.~{Marengo}, M.~T. {Schuster}, J.~L. {Hora},
  R.~G. {Ellis}, J.~R. {Stauffer}, S.~M. {Sonnett}, E.~{Winston}, R.~A.
  {Gutermuth}, S.~T. {Megeath}, D.~E. {Backman}, T.~J. {Henry}, M.~W. {Werner},
  and G.~G. {Fazio}, \emph{\apj} \textbf{654}, 570--579 (2007).

\bibitem{2007ApJ...659.1629L}
K.~L. {Luhman}, K.~N. {Allers}, D.~T. {Jaffe}, M.~C. {Cushing}, K.~A.
  {Williams}, C.~L. {Slesnick}, and W.~D. {Vacca}, \emph{\apj} \textbf{659},
  1629--1636 (2007).

\bibitem{2009ApJ...691.1265L}
K.~L. {Luhman}, E.~E. {Mamajek}, P.~R. {Allen}, A.~A. {Muench}, and D.~P.
  {Finkbeiner}, \emph{\apj} \textbf{691}, 1265--1275 (2009).

\bibitem{2005ApJ...634.1385M}
E.~E. {Mamajek}, \emph{\apj} \textbf{634}, 1385--1394 (2005).

\bibitem{2007ApJ...655..541M}
M.~S. {Marley}, J.~J. {Fortney}, O.~{Hubickyj}, P.~{Bodenheimer}, and J.~J.
  {Lissauer}, \emph{\apj} \textbf{655}, 541--549 (2007).

\bibitem{2008Sci...322.1348M}
C.~{Marois}, B.~{Macintosh}, T.~{Barman}, B.~{Zuckerman}, I.~{Song},
  J.~{Patience}, D.~{Lafreni{\`e}re}, and R.~{Doyon}, \emph{Science}
  \textbf{322}, 1348-- (2008).

\bibitem{2006ApJ...651.1166M}
S.~A. {Metchev}, and L.~A. {Hillenbrand}, \emph{\apj} \textbf{651}, 1166--1176
  (2006).

\bibitem{2005AA...435L..13N}
R.~{Neuh{\"a}user}, E.~W. {Guenther}, G.~{Wuchterl}, M.~{Mugrauer},
  A.~{Bedalov}, and P.~H. {Hauschildt}, \emph{\aap} \textbf{435}, L13--L16
  (2005).

\bibitem{2005astro.ph..9906N}
R.~{Neuh\"auser}, \emph{ArXiv Astrophysics e-prints: astro-ph/0509906}  (2005).

\bibitem{2008AA...484..281N}
R.~{Neuh{\"a}user}, M.~{Mugrauer}, A.~{Seifahrt}, T.~O.~B. {Schmidt}, and
  N.~{Vogt}, \emph{\aap} \textbf{484}, 281--291 (2008).

\bibitem{2008AA...491..311S}
T.~O.~B. {Schmidt}, R.~{Neuh{\"a}user}, A.~{Seifahrt}, N.~{Vogt}, A.~{Bedalov},
  C.~{Helling}, S.~{Witte}, and P.~H. {Hauschildt}, \emph{\aap} \textbf{491},
  311--320 (2008).

\bibitem{2007AA...463..309S}
A.~{Seifahrt}, R.~{Neuh{\"a}user}, and P.~H. {Hauschildt}, \emph{\aap}
  \textbf{463}, 309--313 (2007).

\bibitem{2001ApJ...546..352S}
I.~{Song}, J.-P. {Caillault}, D.~{Barrado y Navascu{\'e}s}, and J.~R.
  {Stauffer}, \emph{\apj} \textbf{546}, 352--357 (2001).

\bibitem{2007ApJ...664.1154S}
K.~G. {Stassun}, R.~D. {Mathieu}, and J.~A. {Valenti}, \emph{\apj}
  \textbf{664}, 1154--1166 (2007).

\end{thebibliography}

\IfFileExists{\jobname.bbl}{}
 {\typeout{}
  \typeout{******************************************}
  \typeout{** Please run "bibtex \jobname" to optain}
  \typeout{** the bibliography and then re-run LaTeX}
  \typeout{** twice to fix the references!}
  \typeout{******************************************}
  \typeout{}
 }


\end{document}